\documentclass[submission,copyright,creativecommons]{eptcs}
\providecommand{\event}{WWV 2013}

\usepackage{amssymb}
\usepackage{graphicx}
\usepackage{times}
\usepackage{latexsym}
\usepackage{amsmath}
\usepackage{stmaryrd}
\usepackage[dvipsnames]{xcolor}
\usepackage{url}
\usepackage{soul}
\usepackage{ulem}
\usepackage[colorinlistoftodos,textsize=footnotesize,textwidth=1.7cm]{todonotes}
\usepackage{cmll}
\usepackage{xspace}
\usepackage{listings}
\usepackage{wasysym}

\newcommand{\code}[1]{{\lstinline[basicstyle=\ttfamily]!#1!}}
\newcommand{\codeTag}[1]{{\lstinline[basicstyle=\ttfamily]!<#1>!}}

\definecolor{mygray}{rgb}{.90,.90,.90}
\definecolor{myDarkGray}{rgb}{.75,.75,.75}
\definecolor{myLightGray}{rgb}{.965,.965,.965}
\definecolor{back}{gray}{1}
\definecolor{rcv}{gray}{0.9}
\definecolor{inv}{gray}{0.9}
\definecolor{ass}{gray}{0.8}

\newcommand{\bgInstanceA}[1]{\colorbox{mygray}{#1}} 
\newcommand{\bgInstanceB}[1]{\colorbox{myDarkGray}{#1}} 

{\begin{trivlist}%
\item[]{\it Proof (sketch).}
}%
{\qed
\end{trivlist}
}

{\begin{trivlist}%
\item[]{\it Proof.}
}%
{\qed
\end{trivlist}
}

\newcommand{\sep}{\quad | \quad}

\newcommand{\pic}{$\pi$-calculus} 

\newcommand{\wsbpel}{\textsc{WS-BPEL}}

\newcommand{\wsdl}{\textsc{WSDL}}

\newcommand{\und}{\_} 

\makeatletter
\def \rightarrowfill{\m@th\mathord{\smash-}\mkern-6mu%
  \cleaders\hbox{$\mkern-2mu\mathord{\smash-}\mkern-2mu$}\hfill
  \mkern-6mu\mathord\rightarrow}
\makeatother

\makeatletter
\def \wrightarrowfill{\m@th\mathord{\smash=}\mkern-6mu%
  \cleaders\hbox{$\mkern-2mu\mathord{\smash=}\mkern-2mu$}\hfill
  \mkern-6mu\mathord\Rightarrow}
\makeatother

\makeatletter
\def \nrightarrowfill{\m@th\mathord{\smash-}\mkern-6mu%
  \cleaders\hbox{$\mkern-2mu\mathord{\smash-}\mkern-2mu$}\hfill
  \mkern-6mu\mathord\nrightarrow}
\makeatother

\def \overstackrel#1#2{\mathrel{\mathop{#1}\limits^{#2}}}
\newcommand{\delimite}[1]{\mbox{$\overstackrel{\belowfill}{\mbox{$\overstackrel{\quad #1 \quad}{\abovefill}$}}$}}
\makeatletter
\def \abovefill{
  {\vrule width0.3mm height 1.8mm depth-0.3mm}
  \leaders\hrule height1.8mm depth-1.5mm\hfill
  {\vrule width0.3mm height 1.8mm depth-0.3mm}}
\makeatother

\makeatletter
\def \belowfill{
  {\vrule width0.3mm height1.5mm}
  \leaders\hrule height0.3mm\hfill
  {\vrule width0.3mm height1.5mm}}
\makeatother

\newcommand{\cows}{\textsc{COWS}} 

\newcommand{\partner}{p} 
\newcommand{\op}{o} 
\newcommand{\expr}{\epsilon} 
\newcommand{\epc}[2]{#1 \!\stackrel{{}_\bullet}{} \!#2} 
\newcommand{\var}{x} 
\newcommand{\varBis}{y} 
\newcommand{\name}{n} 
\newcommand{\nameBis}{m} 
\newcommand{\val}{v} 
\newcommand{\xorn}{u} 
\newcommand{\xorv}{w} 
\newcommand{\guard}{g} 

\newcommand{\out}[3]{\epc{#1}{#2} ! #3} 

\newcommand{\spar}{\mid} 

\newcommand{\scope}[1]{[#1]\,} 

\newcommand{\rep}{\ast\,} 

\newcommand{\nil}{\mathbf{0}} 

\newcommand{\inp}[3]{\epc{#1}{#2} ? #3} 

\newcommand{\arr}[1]{\langle #1 \rangle} 

\newcommand{\define}{\triangleq} 

\newdimen\proofrulebreadth \proofrulebreadth=.04em
\newdimen\proofdotseparation \proofdotseparation=1.25ex
\newdimen\proofrulebaseline \proofrulebaseline=2ex
\newcount\proofdotnumber \proofdotnumber=3
\let\then\relax
\def\hfi{\hskip0pt plus.0001fil}
\mathchardef\squigto="3A3B
\newif\ifinsideprooftree\insideprooftreefalse
\newif\ifonleftofproofrule\onleftofproofrulefalse
\newif\ifproofdots\proofdotsfalse
\newif\ifdoubleproof\doubleprooffalse
\let\wereinproofbit\relax
\newdimen\shortenproofleft
\newdimen\shortenproofright
\newdimen\proofbelowshift
\newbox\proofabove
\newbox\proofbelow
\newbox\proofrulename
\def\shiftproofbelow{\let\next\relax\afterassignment\setshiftproofbelow\dimen0 }
\def\shiftproofbelowneg{\def\next{\multiply\dimen0 by-1 }%
\afterassignment\setshiftproofbelow\dimen0 }
\def\setshiftproofbelow{\next\proofbelowshift=\dimen0 }
\def\setproofrulebreadth{\proofrulebreadth}

\def\prooftree{
\ifnum  \lastpenalty=1 \then   \unpenalty \else
\onleftofproofrulefalse \fi
\ifonleftofproofrule \else   \ifinsideprooftree
        \then   \hskip.5em plus1fil
        \fi
\fi
\bgroup
\setbox\proofbelow=\hbox{}\setbox\proofrulename=\hbox{}%
\let\justifies\proofover\let\leadsto\proofoverdots\let\Justifies\proofoverdbl
\let\using\proofusing\let\[\prooftree
\ifinsideprooftree\let\]\endprooftree\fi
\proofdotsfalse\doubleprooffalse
\let\thickness\setproofrulebreadth
\let\shiftright\shiftproofbelow \let\shift\shiftproofbelow
\let\shiftleft\shiftproofbelowneg
\let\ifwasinsideprooftree\ifinsideprooftree
\insideprooftreetrue
\setbox\proofabove=\hbox\bgroup$\displaystyle 
\let\wereinproofbit\prooftree
\shortenproofleft=0pt \shortenproofright=0pt \proofbelowshift=0pt
\onleftofproofruletrue\penalty1 }

\def\eproofbit{
\ifx    \wereinproofbit\prooftree \then   \ifcase \lastpenalty
        \then   \shortenproofright=0pt  
        \or     \unpenalty\hfil         
        \or     \unpenalty\unskip       
        \else   \shortenproofright=0pt  
        \fi
\fi
\global\dimen0=\shortenproofleft \global\dimen1=\shortenproofright
\global\dimen2=\proofrulebreadth \global\dimen3=\proofbelowshift
\global\dimen4=\proofdotseparation
$\egroup  
\shortenproofleft=\dimen0 \shortenproofright=\dimen1
\proofrulebreadth=\dimen2 \proofbelowshift=\dimen3
\proofdotseparation=\dimen4
}

\def\proofover{
\eproofbit 
\setbox\proofbelow=\hbox\bgroup 
\let\wereinproofbit\proofover
$\displaystyle
}%
\def\proofoverdbl{
\eproofbit 
\doubleprooftrue
\setbox\proofbelow=\hbox\bgroup 
\let\wereinproofbit\proofoverdbl
$\displaystyle
}%
\def\proofoverdots{
\eproofbit 
\proofdotstrue
\setbox\proofbelow=\hbox\bgroup 
\let\wereinproofbit\proofoverdots
$\displaystyle
}%
\def\proofusing{
\eproofbit 
\setbox\proofrulename=\hbox\bgroup 
\let\wereinproofbit\proofusing
\kern0.3em$ }

\def\endprooftree{
\eproofbit 
  \dimen5 =0pt
\dimen0=\wd\proofabove \advance\dimen0-\shortenproofleft
\advance\dimen0-\shortenproofright
\dimen1=.5\dimen0 \advance\dimen1-.5\wd\proofbelow \dimen4=\dimen1
\advance\dimen1\proofbelowshift \advance\dimen4-\proofbelowshift
\ifdim  \dimen1<0pt \then   \advance\shortenproofleft\dimen1
        \advance\dimen0-\dimen1
        \dimen1=0pt
        \ifdim  \shortenproofleft<0pt
        \then   \setbox\proofabove=\hbox{%
                        \kern-\shortenproofleft\unhbox\proofabove}%
                \shortenproofleft=0pt
        \fi
\fi
\ifdim  \dimen4<0pt \then   \advance\shortenproofright\dimen4
        \advance\dimen0-\dimen4
        \dimen4=0pt
\fi
\ifdim  \shortenproofright<\wd\proofrulename \then
\shortenproofright=\wd\proofrulename \fi
\dimen2=\shortenproofleft \advance\dimen2 by\dimen1
\dimen3=\shortenproofright\advance\dimen3 by\dimen4
\ifproofdots \then
        \dimen6=\shortenproofleft \advance\dimen6 .5\dimen0
        \setbox1=\vbox to\proofdotseparation{\vss\hbox{$\cdot$}\vss}
        \setbox0=\hbox{%
                \kern\dimen6
                $\vcenter to\proofdotnumber\proofdotseparation
                        {\leaders\box1\vfill}$%
                \unhbox\proofrulename}%
\else   \dimen6=\fontdimen22\the\textfont2 
        \dimen7=\dimen6
        \advance\dimen6by.5\proofrulebreadth
        \advance\dimen7by-.5\proofrulebreadth
        \setbox0=\hbox{%
                \kern\shortenproofleft
                \ifdoubleproof
                \then   \hbox to\dimen0{%
                        $\mathsurround0pt\mathord=\mkern-6mu%
                        \cleaders\hbox{$\mkern-2mu=\mkern-2mu$}\hfill
                        \mkern-6mu\mathord=$}%
                \else   \vrule height\dimen6 depth-\dimen7 width\dimen0
                \fi
                \unhbox\proofrulename}%
        \ht0=\dimen6 \dp0=-\dimen7
\fi
\let\doll\relax
\ifwasinsideprooftree \then   \let\VBOX\vbox \else
\ifmmode\else$\let\doll=$\fi
        \let\VBOX\vcenter
\fi
\VBOX   {\baselineskip\proofrulebaseline \lineskip.2ex
        \expandafter\lineskiplimit\ifproofdots0ex\else-0.6ex\fi
        \hbox   spread\dimen5   {\hfi\unhbox\proofabove\hfi}%
        \hbox{\box0}%
        \hbox   {\kern\dimen2 \box\proofbelow}}\doll%
\global\dimen2=\dimen2 \global\dimen3=\dimen3
\egroup 
\ifonleftofproofrule \then   \shortenproofleft=\dimen2 \fi
\shortenproofright=\dimen3
\onleftofproofrulefalse \ifinsideprooftree \then   \hskip.5em plus
1fil \penalty2 \fi }

\makeatletter
\def \Rightarrowfill{\m@th\mathord{\smash=}\mkern-6mu%
  \cleaders\hbox{$\mkern-2mu\mathord{\smash=}\mkern-2mu$}\hfill
  \mkern-6mu\mathord\Rightarrow}
\makeatother

\title{
  Blind-date Conversation Joining
  \thanks{This work has been partially sponsored by the EU project ASCENS (257414).}
}

\author{
Luca Cesari$^{1,2}$ 
\institute{$^1$Universit\`a di Pisa, Italy}
\email{cesari@di.unipi.it} 
\and
Rosario Pugliese$^2$
\institute{$^2$Universit\`a degli Studi di Firenze, Italy}
\email{luca.cesari@unifi.it}
\email{rosario.pugliese@unifi.it}
\and
Francesco Tiezzi$^3$
\institute{$^3$IMT Advanced Studies Lucca, Italy}
\email{francesco.tiezzi@imtlucca.it} 
}

\def\titlerunning{Blind-date Conversation Joining}
\def\authorrunning{L. Cesari, R. Pugliese \& F. Tiezzi}

\begin{document}

\maketitle

\begin{abstract}
  We focus on a form of joining conversations among multiple parties in
  service-oriented applications where a client may asynchronously join an
  existing conversation without need to know in advance any information
  about it. More specifically, we show how the correlation mechanism
  provided by orchestration languages enables a form of conversation
  joining that is completely transparent to clients and that we call
  `blind-date joining'. We provide an implementation of this strategy by
  using the standard orchestration language \wsbpel. We then present its
  formal semantics by resorting to \cows, a process calculus specifically
  designed for modelling service-oriented applications. We illustrate our
  approach by means of a simple, but realistic, case study from the online
  games domain.
\end{abstract}

\section{Introduction}
\label{sec:intro}

The increasing success of e-business, e-learning, e-government, and other
similar emerging models, has led the World Wide Web, initially thought of
as a system for human use, to evolve towards an architecture for supporting
automated use. A new computing paradigm has emerged, called
Service-Oriented Computing (SOC), that advocates the use of loosely-coupled
\emph{services}. These are autonomous, platform-independent, computational
entities that can be described, published, discovered, and assembled as the
basic blocks for building interoperable and evolvable systems and
applications. Currently, the most successful instantiation of the SOC
paradigm are \emph{Web Services}, i.e. sets of operations that can be
published, located and invoked through the Web via XML messages complying
with given standard formats.

In SOC, service definitions are used as templates for creating service
instances that deliver application functionalities to either end-user
applications or other instances. Upon service invocation, differently from
what usually happens in traditional client-server paradigms, the caller
(i.e., a service client) and the callee (i.e., a service provider) can
engage in a \emph{conversation} during which they exchange the information
needed to complete all the activities related to the specific service. For
instance, a client of an airplane ticket reservation service usually
interacts several times with the service before selecting the specific
flight to be reserved. Although initially established between a caller and
callee, a conversation can dynamically accommodate and dismiss
participants. Therefore, a conversation is typically a loosely-coupled,
multiparty interaction among a (possibly dynamically varying) number of
participants.

The loosely-coupled nature of SOC implies that, from a technological point
of view, the connection between communicating partners cannot be assumed to
persist for the duration of a whole conversation. Even the execution of a
simple request-response message exchange pattern provides no built-in means
of automatically associating the response message with the original
request. It is up to each single message to provide a form of context thus
enabling partners to associate the message with others. This is achieved by
including values in the message which, once located, can be used to
correlate the message with others logically forming the same stateful
conversation. The link among partners is thus determined by so called
\emph{correlation values}: only messages containing the `right' correlation
values are processed by the partner.

Message correlation is an essential part of messaging within SOC as it
enables the persistence of activities' context and state across multiple
message exchanges while preserving service statelessness and autonomy, and
the loosely-coupled nature of SOC systems. It is thus at the basis of Web
Services interaction which is implemented on top of stateless Internet
protocols, such as the transfer protocol HTTP. For example, the Internet
Cookies used by web sites in order to relate an HTTP request to a user
profile, thus enabling to return a customized HTML file to the user, are
correlation information. Besides being useful to implement stateful
communication on top of stateless protocols, correlation is also a flexible
and user programmable mechanism for managing loosely-coupled, multiparty
conversations. Indeed, correlation data can be communicated to other
partners in order to allow them to join a conversation or to delegate the
task of carrying out an ongoing conversation.

In this paper, we show another evidence of the flexibility of the
correlation mechanism that involves creation of and joining conversations.
More specifically, we demonstrate how correlation allows a partner to
asynchronously join an existing conversation without need to know in
advance any information about the conversation itself, such as e.g. its
identifier or the other participants. Since this particular kind of
conversation joining is completely transparent to participants we call it
\emph{blind-date joining}. It can be used in various domains like
e-commerce, events organization, bonus payments, gift lists, etc. In the
context of e-commerce, for instance, a social shopping provider (e.g.,
Groupon~\cite{GROUPON}) can activate deals only if a certain number of
buyers adhere. In this scenario, the blind-date joining can be used to
activate these deals. As another example, blind-date joining can be used to
organize an event only if a given number of participants is reached.
Specifically, the organizers can wait for the right amount of participants
and, when this is reached, they send the invitation with the location
details to each of them.

We illustrate the blind-date joining strategy by exploiting a simple but
realistic case study from the online games domain. We then implement the
case study via a well-established orchestration language for web services,
i.e. the OASIS standard \wsbpel~\cite{WSBPEL}. To better clarify and
formally present the semantics of the blind-date joining strategy we also
introduce a specification of the case study, and its step-by-step temporal
evolution, using the process calculus \cows~\cite{LPT07:ESOP,COWS_JAL}, a
formalism specifically designed for specifying and combining SOC
applications, while modelling their dynamic behaviour.

The rest of the paper is structured as follows.
Section~\ref{sec:background} surveys syntax and semantics of \wsbpel\ and
\cows. Section~\ref{sec:caseStudy} presents how blind-date conversation can
be expressed in \wsbpel\ by implementing a case study from the online games
domain. Section~\ref{sec:caseStudycows} formally describes, by using \cows,
the creation and joining phase of blind-date conversations. Finally,
Section~\ref{sec:conclusion} concludes the paper by also reviewing more
closely related work.

\section{A glimpse of \wsbpel\ and \cows}
\label{sec:background}

This section presents a survey of \wsbpel\ and \cows. The overview of
\wsbpel\ gives a high-level description of the aspects captured by its
linguistic constructs. Due to lack of space, the overview of \cows\ gives
only a glimpse of its semantics, a full account of which can be found
in~\cite{COWS_JAL}.

\subsection{An overview of \wsbpel}
\label{sec:wsbpel}

\wsbpel~\cite{WSBPEL} is essentially a linguistic layer on top of \wsdl\
for describing the structural aspects of Web Services `orchestration', i.e.
the process of combining and coordinating different Web Services to obtain
a new, customised service. In practice, and briefly, \wsdl\ \cite{WSDL} is
a W3C standard that permits to express the functionalities offered and
required by web services by defining, akin object interfaces in
Object-Oriented Programming, the structure of input and output messages of
operations.

In \wsbpel, the logic of interaction between a service and its environment
is described in terms of structured patterns of communication actions
composed by means of control flow constructs that enable the representation
of complex structures. For the specification of orchestration, \wsbpel\
provides many different activities that are distinguished between
\emph{basic activities} and \emph{structured activities}. Orchestration
exploits state information that is stored in variables and is managed
through message correlation. In fact, when messages are sent/received, the
value of their parameters is stored in variables. Likewise block structured
languages, the scope of variables extends to the whole immediately
enclosing \codeTag{scope}, or \codeTag{process}, activity (whose meaning is
clarified below).

The basic activities are:
\codeTag{invoke}, to invoke an operation offered by a Web Service;
\codeTag{receive}, to wait for an invocation to arrive;
\codeTag{reply}, to send a message in reply to a previously received invocation;
\codeTag{wait}, to delay execution for some amount of time;
\codeTag{assign}, to update the values of variables with new data;
\codeTag{throw}, to signal internal faults;
\codeTag{exit}, to immediately end a service instance;
\codeTag{empty}, to do nothing;
\codeTag{compensate} and \codeTag{compensateScope}, to invoke compensation handlers;
\codeTag{rethrow}, to propagate faults;
\codeTag{validate}, to validate variables; and
\codeTag{extensionActivity}, to add new activity types.
Notably, \codeTag{reply} can be combined with \codeTag{receive}
to model synchronous request-response interactions.

The structured activities describe the control flow logic of a business
process by composing basic and/or structured activities recursively. The
structured activities are: \codeTag{sequence}, to execute activities
sequentially; \codeTag{if}, to execute activities conditionally;
\codeTag{while} and \codeTag{repeatUntil}, to repetitively execute
activities; \codeTag{flow}, to execute activities in parallel;
\codeTag{pick}, to execute activities selectively; \codeTag{forEach}, to
(sequentially or in parallel) execute multiple activities; and
\codeTag{scope}, to associate handlers for exceptional events to a primary
activity.  Activities within a \codeTag{flow} can be further synchronised
by means of \emph{flow links}.  These are conditional transitions
connecting activities to form directed acyclic graphs and are such that a
target activity may only start when all its source activities have
completed and the condition on the incoming flow links evaluates to true.

The handlers within a \codeTag{scope} can be of four different kinds:
\codeTag{faultHandler}, to provide the activities in response to faults
occurring during execution of the primary activity;
\codeTag{compensationHandler}, to provide the activities to compensate the
successfully executed primary activity; \codeTag{terminationHandler}, to
control the forced termination of the primary activity; and
\codeTag{eventHandler}, to process message or timeout events occurring
during execution of the primary activity. If a fault occurs during
execution of a primary activity, the control is transferred to the
corresponding fault handler and all currently running activities inside the
scope are interrupted immediately without involving any fault/compensation
handling behaviour. If another fault occurs during a fault/compensation
handling, then it is re-thrown, possibly, to the immediately enclosing
scope. Compensation handlers attempt to reverse the effects of previously
successfully completed primary activities (scopes) and have been introduced
to support Long-Running (Business) Transactions (LRTs). Compensation can
only be invoked from within fault or compensation handlers starting the
compensation either of a specific inner (completed) scope, or of all inner
completed scopes in the reverse order of completion. The latter alternative
is also called the \emph{default} compensation behaviour. Invoking a
compensation handler that is unavailable is equivalent to perform an empty
activity.

A \wsbpel\ program, also called \emph{(business) process}, is a
\codeTag{process}, that is a sort of \codeTag{scope} without compensation
and termination handlers.

\wsbpel\ uses the basic notion of \emph{partner link} to directly model
peer-to-peer relationships between services. Such a relationship is
expressed at the \wsdl\ level by specifying the roles played by each of the
services in the interaction. This information, however, does not suffices
to deliver messages to a service. Indeed, since multiple instances of the
same service can be simultaneously active because service operations can be
independently invoked by several clients, messages need to be delivered not
only to the correct partner, but also to the correct instance of the
service that the partner provides. To achieve this, \wsbpel\ relies on the
business data exchanged rather than on specific mechanisms, such as
WS-Addressing~\cite{WSADDRESSING} or low-level methods based on SOAP
headers. In fact, \wsbpel\ exploits \emph{correlation sets}, namely sets of
\emph{correlation variables} (called \emph{properties}), to declare the
parts of a message that can be used to identify an instance. This way, a
message can be delivered to the correct instance on the basis of the values
associated to the correlation variables, independently of any routing
mechanism.

\subsection{An overview of \cows}
\label{sec:cows}

\cows\ \cite{LPT07:ESOP,COWS_JAL} is a formalism for modelling (and
analysing) service-oriented applications. It provides a novel combination
of constructs and features borrowed from well-known process calculi such as
non-binding receiving activities, asynchronous communication, polyadic
synchronization, pattern matching, protection, and delimited receiving and
killing activities. As a consequence of its careful design, the calculus
makes it easy to model many important aspects of service orchestrations
\emph{\`a la} \wsbpel, such as service instances with shared state,
services playing more than one partner role, stateful conversations made by
several correlated service interactions, and long-running transactions. For
the sake of simplicity, we present here a fragment of \cows\ (called
$\mu\cows$ in \cite{COWS_JAL}) without linguistic constructs for dealing
with `forced termination', since such primitives are not used in this work.

The syntax of \cows\ is presented in Table~\ref{tab:syntaxCOWSmm}. We use
two countable disjoint sets: the set of \emph{values} (ranged over by
$\val$, $\val'$, \ldots) and the set of `write once' \emph{variables}
(ranged over by $\var$, $\varBis$,~\ldots). The set of values is left
unspecified; however, we assume that it includes the set of \emph{names}
(ranged over by $\name$, $\nameBis$, $\partner$, $\op$, \ldots) mainly used
to represent partners and operations. We also use a set of
\emph{expressions} (ranged over by $\expr$), whose exact syntax is
deliberately omitted; we just assume that expressions contain, at least,
values and variables. As a matter of notation, $\xorv$ ranges over values
and variables and $\xorn$ ranges over names and variables. Notation $\bar
\cdot$ stands for tuples, e.g. $\bar \var$  means $\arr{x_1,\ldots,x_n}$
(with $n \geq 0$) where variables in the same tuple are pairwise distinct.
We will omit trailing occurrences of $\nil$, writing e.g.
$\inp{\partner}{\op}{\bar \xorv}$ instead of $\inp{\partner}{\op}{\bar
\xorv}.\nil$, and write $\scope{\arr{\xorn_1,\ldots,\xorn_n}} s$ in place
of $\scope{\xorn_1}\ldots\scope{\xorn_n} s$. We will write $I \define s$ to
assign a name $I$ to the term $s$.

\emph{Services} are structured activities built from basic activities, i.e.
the empty activity $\nil$, the invoke activity $\out{\und}{\und}{\und}$ and
the receive activity $\inp{\und}{\und}{\und}$\,, by means of prefixing
$\und\, . \und$\,, choice $\und + \und$\,, parallel composition $\und \spar
\und$\,, delimitation $\scope{\und} \und$ and replication $\rep \und$\,. We
adopt the following conventions about the operators precedence: monadic
operators bind more tightly than parallel composition, and prefixing more
tightly than choice.

\emph{Invoke} and \emph{receive} are the communication activities, which
permit invoking an operation offered by a service and waiting for an
invocation to arrive, respectively. Besides output and input parameters,
both activities indicate an \emph{endpoint}, i.e. a pair composed of a
partner name $\partner$  and an operation name $\op$, through which
communication should occur. An endpoint $\epc{\partner}{\op}$ can be
interpreted as a specific implementation of operation $\op$ provided by the
service identified by the logic name $\partner$. An invoke
$\out{\partner}{\op}{\arr{\expr_{1},\ldots,\expr_{n}}}$ can proceed as soon
as all expression arguments $\expr_{1}$, \ldots, $\expr_{n}$ can be
successfully evaluated to values. A receive
$\inp{\partner}{\op}{\arr{\xorv_{1},\ldots,\xorv_{n}}}.s$ offers an
invocable operation $\op$ along a given partner name $\partner$ and
thereafter (due to the \emph{prefixing} operator) the service continues as
$s$. An inter-service communication between these two activities takes
place when the tuple of values $\arr{\val_{1},\ldots,\val_{n}}$, resulting
from the evaluation of the invoke argument, matches the template
$\arr{\xorv_{1},\ldots,\xorv_{n}}$ argument of the receive. This causes a
substitution of the variables in the receive template (within the scope of
variables declarations) with the corresponding values produced by the
invoke. Partner and operation names can be exchanged in communication, thus
enabling many different interaction patterns among service instances.
However, dynamically received names cannot form the endpoints used to
receive further invocations. Indeed, endpoints of receive activities are
identified statically because their syntax only allows using names and not
variables.

The \emph{empty} activity does nothing, while \emph{choice} permits
selecting for execution one between two alternative receives.

Execution of \emph{parallel} services is interleaved. However, if more
matching receives are ready to process a given invoke, only one of the
receives that generate a substitution with smallest domain (see
\cite{COWS_JAL} for further details) is allowed to progress (namely, this
receive has an higher priority to proceed than the others). This mechanism
permits to model the precedence of a service instance over the
corresponding service specification when both can process the same request
(see also Section~\ref{sec:caseStudycows}).

\begin{table}[t]
\begin{center}
\delimite{
\begin{tabular}{@{\hspace{-.2cm}}l@{\hspace{.2cm}}c@{\hspace{.2cm}}l@{\hspace{.3cm}}l@{\hspace{.3cm}}|@{\hspace{.3cm}}l@{\hspace{.2cm}}c@{\hspace{.2cm}}l@{\hspace{.3cm}}l@{\hspace{-.1cm}}}
\\[-.7cm]
\multicolumn{4}{@{\hspace{-.2cm}}l}{ \emph{Expressions:} $\expr$, $\expr'$, \ldots} &
\\
\multicolumn{4}{@{\hspace{-.2cm}}l}{ \qquad\emph{Values:} $\val$, $\val'$, \ldots}
& \multicolumn{4}{l}{\hspace*{-.21cm}\emph{Variables/Values:}
$\xorv$, $\xorv'$, \ldots}
\\
\multicolumn{4}{@{\hspace{-.2cm}}l}{ \qquad\qquad\emph{Names:} $\name$, $\nameBis$,
\ldots} & \multicolumn{4}{l}{\hspace*{-.21cm}\emph{Variables/Names:} $\xorn$, $\xorn'$, \ldots}
\\
\multicolumn{4}{@{\hspace{-.2cm}}l}{ \qquad\qquad\qquad\emph{Partners:} $\partner$,
$\partner'$, \ldots}
\\
\multicolumn{4}{@{\hspace{-.2cm}}l}{ \qquad\qquad\qquad\emph{Operations:} $\op$,
$\op'$, \ldots}  &
\\
\multicolumn{4}{@{\hspace{-.2cm}}l}{ \qquad\emph{Variables:} $\var$, $\varBis$,
\ldots}  &
\\
\\[-.4cm]
\hline
\\[-.4cm]
\multicolumn{3}{@{\hspace{-.2cm}}l}{\emph{Services:}} & &
\multicolumn{4}{l}{\hspace*{-.21cm}\emph{Receive-guarded choice:}}\\
$s$ & ::=      & & &
$\guard$ & ::=      & &  \\
    & $    $   & $\out{\xorn}{\xorn'}{\bar \expr}$ & \quad (invoke)
&
    & $ $      & $\nil$ & \quad (empty) \\
    & $\sep$   & $\guard$ & \quad (receive-guarded choice)
&
    & $\sep$   & $\inp{\partner}{\op}{\bar \xorv}.s$ & \quad (receive) \\
    & $\sep$   & $s \spar s$ & \quad (parallel composition)
&
    & $\sep$   & $\guard + \guard$ & \quad (choice)\\
    & $\sep$   & $\scope{\xorn} s$ & \quad (delimitation) &&&&\\
    & $\sep$   & $\rep s$ & \quad (replication) &&&&\\[-.1cm]
\end{tabular}
}
\end{center}
\vspace*{-0.7cm} \caption{\cows\ syntax} \label{tab:syntaxCOWSmm}
\vspace*{-.4cm}
\end{table}

\emph{Delimitation} is the only binding construct: $\scope{\xorn} s$ binds
the element $\xorn$ in the scope $s$. It is used for two different
purposes: to regulate the range of application of substitutions produced by
communication, if the delimited element is a variable, and to generate
fresh names, if the delimited element is a name. 

Finally, the \emph{replication} construct $\rep s$ permits to spawn in
parallel as many copies of $s$ as necessary.  This, for example, is
exploited to implement recursive behaviours and to model business process
definitions, which can create multiple instances to serve several requests
simultaneously.

\section{Blind-date conversations in \wsbpel}
\label{sec:caseStudy}

We present here a (web) service capable of arranging matches of
4-players\footnote{
  It is worth noticing that the blind-date conversation
  approach works as well with a number of players not fixed a priori. Of
  course, this would require using some extra activities.  Therefore, to
  better highlight the specificities of the proposed approach, we have
  preferred to keep the example as simple as possible and, hence, we have
  fixed the number of players to four.
}
online (card) games, such as e.g.  \emph{burraco} or
\emph{canasta}. To create or join a match, a player has only to indicate
the kind of game he/she would like to play and his/her endpoint reference
(i.e., his/her address). Thus, players do not need to know in advance any
further information, such as the identifier of the table or the identifiers
of other players.  Moreover, players do not either know if a new match is
created as a consequence of their request to play, or if they join an
already existing match. Therefore, the arrangement of tables is
\emph{completely transparent} to players.

\begin{figure}[t]
  \begin{center}
    \includegraphics[scale=.55]{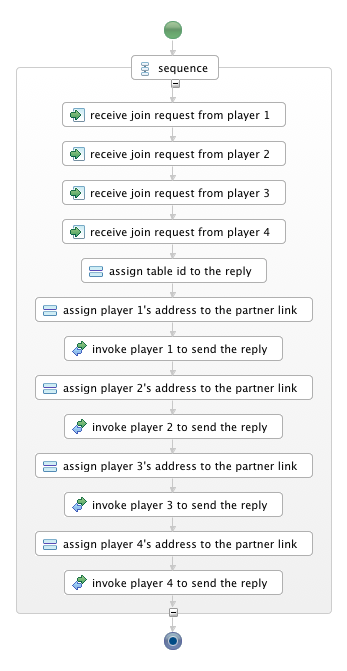}
  \end{center}
  \vspace*{-1cm}
  \caption{Graphical representation of the \wsbpel\ process \code{TableManager}}
  \label{fig:process}
  \vspace*{-.3cm}
\end{figure}

Figure~\ref{fig:process} shows a graphical representation\footnote{
  The representation has been created by using Eclipse
  BPEL Designer (\url{http://www.eclipse.org/bpel/}).
}
We report below the code of the process where, to
simplify the reading of the code, we have omitted irrelevant details and
highlighted the basic activities \codeTag{receive}, \codeTag{invoke} and
\codeTag{assign} by means of a grey background.

\begin{footnotesize}
\begin{center}

\lstset{ basicstyle=\ttfamily}
\begin{lstlisting}
<process name="TableManager" ... >
<partnerLinks>
  <partnerLink name="tableManager"
               myRole="table" partnerRole="player"
               initializePartnerRole="no"
               partnerLinkType="tableManager:table"/>
</partnerLinks>
<variables>
  <variable messageType="tableManager:request" name="Player1"/>
  <variable messageType="tableManager:request" name="Player2"/>
  <variable messageType="tableManager:request" name="Player3"/>
  <variable messageType="tableManager:request" name="Player4"/>
  <variable messageType="tableManager:reply" name="PlayersReply"/>
</variables>
<correlationSets>
  <correlationSet name="GameCorrelationSet"
                  properties="tableManager:gameNameProperty"/>
</correlationSets>
<sequence>
\end{lstlisting}

\vspace{-3ex}
\lstset{ basicstyle=\ttfamily, backgroundcolor=\color{rcv}}
\begin{lstlisting}
  <receive partnerLink="tableManager"
           operation="join"
           variable="Player1"
           createInstance="yes">
        <correlations>
          <correlation initiate="yes" set="GameCorrelationSet"/>
        </correlations>
  </receive>
\end{lstlisting}

\vspace{-3ex}
\lstset{ basicstyle=\ttfamily, backgroundcolor=\color{rcv}}
\begin{lstlisting}
  <receive partnerLink="tableManager"
           operation="join"
           variable="Player2">
        <correlations>
          <correlation initiate="no" set="GameCorrelationSet"/>
        </correlations>
  </receive>
\end{lstlisting}

\vspace{-3ex}
\lstset{ basicstyle=\ttfamily, backgroundcolor=\color{rcv}}
\begin{lstlisting}
  <receive partnerLink="tableManager"
           operation="join"
           variable="Player3">
        <correlations>
          <correlation initiate="no" set="GameCorrelationSet"/>
        </correlations>
  </receive>
\end{lstlisting}

\vspace{-3ex}
\lstset{ basicstyle=\ttfamily, backgroundcolor=\color{rcv}}
\begin{lstlisting}
  <receive partnerLink="tableManager"
           operation="join"
           variable="Player4">
        <correlations>
          <correlation initiate="no" set="GameCorrelationSet"/>
        </correlations>
  </receive>
\end{lstlisting}

\vspace{-3ex}
\lstset{ basicstyle=\ttfamily, backgroundcolor=\color{ass}}
\begin{lstlisting}
  <assign>
    ...
    <copy ...>
      <from>getTableID()</from>
      <to>$PlayersReply.payload/tableManager:tableID</to>
    </copy>
  </assign>
\end{lstlisting}

\vspace{-3ex}
\lstset{ basicstyle=\ttfamily, backgroundcolor=\color{ass}}
\begin{lstlisting}
  <assign>
    <copy>
      <from>$Player1.payload/tableManager:playerAddress</from>
      <to partnerLink="tableManager"/>
    </copy>
  </assign>
\end{lstlisting}

\vspace{-3ex}
\lstset{ basicstyle=\ttfamily, backgroundcolor=\color{inv}}
\begin{lstlisting}
  <invoke inputVariable="PlayersReply"
          operation="start"
          partnerLink="tableManager"/>
\end{lstlisting}

\vspace{-3ex}
\lstset{ basicstyle=\ttfamily, backgroundcolor=\color{ass}}
\begin{lstlisting}
  <assign>
    <copy>
      <from>$Player2.payload/tableManager:playerAddress</from>
      <to partnerLink="tableManager"/>
    </copy>
  </assign>
\end{lstlisting}

\vspace{-3ex}
\lstset{ basicstyle=\ttfamily, backgroundcolor=\color{inv}}
\begin{lstlisting}
  <invoke inputVariable="PlayersReply"
          operation="start"
          partnerLink="tableManager"/>
\end{lstlisting}

\vspace{-3ex}
\lstset{ basicstyle=\ttfamily, backgroundcolor=\color{ass}}
\begin{lstlisting}
  <assign>
    <copy>
      <from>$Player3.payload/tableManager:playerAddress</from>
      <to partnerLink="tableManager"/>
    </copy>
  </assign>
\end{lstlisting}

\vspace{-3ex}
\lstset{ basicstyle=\ttfamily, backgroundcolor=\color{inv}}
\begin{lstlisting}
  <invoke inputVariable="PlayersReply"
          operation="start"
          partnerLink="tableManager"/>
\end{lstlisting}

\vspace{-3ex}
\lstset{ basicstyle=\ttfamily, backgroundcolor=\color{ass}}
\begin{lstlisting}
  <assign>
    <copy>
      <from>$Player4.payload/tableManager:playerAddress</from>
      <to partnerLink="tableManager"/>
    </copy>
  </assign>
\end{lstlisting}

\vspace{-3ex}
\lstset{basicstyle=\ttfamily, backgroundcolor=\color{inv}}
\begin{lstlisting}
  <invoke inputVariable="PlayersReply"
          operation="start"
          partnerLink="tableManager"/>
\end{lstlisting}

\vspace{-3ex}
\lstset{basicstyle=\ttfamily, backgroundcolor=\color{back}}
\begin{lstlisting}
  </sequence>
</process>
\end{lstlisting}

\end{center}
\end{footnotesize}
The process uses only one partner link, namely \code{tableManager}, that
provides two roles: \code{table} and \code{player}. The former is used by
the process to receive the players' requests, while the latter is used by
players to receive the table identifier. Five variables are used for
storing data of the exchanged messages: one for each player request and one
for the manager response. Notably, the used message style\footnote{ 
  The SOAP message style configuration is specified in the
  \emph{binding} section of the \wsdl\ document associated to the \wsbpel\
  process. We have preferred to use the document-style rather than the
  RPC-style, because the former minimizes coupling between the interacting
  parties.
}
is \emph{document}, thus messages are formed by a single part, called
\code{payload}, that contains all message data. Therefore, we use XPath
expressions of the form \code{$VariableName.payload/Path} to extract or
store data in message variables.

The process starts with a \codeTag{receive} activity waiting for a message
from a player; the message contains a request (stored in the variable
\code{Player1}) to participate to a match of a given game.  Whenever
prompted by a player's request, the process creates an instance (see the
option \code{createInstance="yes"} in the first \codeTag{receive}),
corresponding to a new (virtual) card-table of the game specified by the
player, and is immediately ready to concurrently serve other requests.
Service instances are indeed the \wsbpel\ counterpart of inter-service
conversations.  In order to deliver each request to an existing instance
corresponding to a table of the requested game (if there exists one), the
name of the game is used as correlation datum. Thus, each \codeTag{receive}
activity specifies the correlation set \code{GameCorrelationSet}, which is
instantiated by the initial \codeTag{receive}, in order to receive only
requests for the same game indicated by the first request. The
\codeTag{property} defining the correlation set is declared in the WSDL
document associated to the process as follows:

\begin{footnotesize}
\begin{center}
\lstset{basicstyle=\ttfamily}
\begin{lstlisting}
<prop:property name="gameNameProperty" type="xs:string"/>

<prop:propertyAlias messageType="this:request" part="payload"
                    propertyName="this:gameNameProperty">
  <prop:query queryLanguage=...>//this:RequestElement/this:gameName</prop:query>
</prop:propertyAlias>
\end{lstlisting}
\end{center}
\end{footnotesize}
A \codeTag{property} specifies an element of a correlation set and relies
on one (or more) \lstinline[basicstyle=\ttfamily]!<property-!
\lstinline[basicstyle=\ttfamily]!Alias>!  to identify correlation values
inside messages. In our specification, the \codeTag{propertyAlias} extracts
from \codeTag{request} messages the needed element by using an XPath
\codeTag{query}. Then, the correlation set \code{GameCorrelationSet} is
defined by the property \code{gameNameProperty} that identifies the string
element \code{gameName} of the messages sent by players.

Once the initial \codeTag{receive} is executed and an instance is created,
other three \codeTag{receive} activities are sequentially performed by such
instance, in order to complete the card-table for the new match.  Notice
that the correlation mechanism ensures that only players that want to play
the same game are put together in a table.

When four players join a conversation for a new match (which, in \wsbpel,
corresponds to a process instance), a unique table identifier is generated,
by means of the custom XPath expression \code{getTableID()}, and inserted
into the variable \code{PlayersReply}.  This variable contains the message
that will be sent back to each player via four \codeTag{invoke} activities
using the \code{tableManager} partner link.
Before every \codeTag{invoke}, an activity \codeTag{assign} is executed to
extract (by means of an XPath expression) the endpoint reference of the
player, contained within the player's request, and to store it into the
\code{partnerRole} of the \code{tableManager} partner link.
These assignments allow the process to properly reply in an asynchronous
way to the players.

Now, the new table is arranged and, therefore, the players can start to
play by using the received table identifier and by interacting with another
service dedicated to this purpose (which, of course, is out of the scope of
this work).

\paragraph{Experimenting with the \wsbpel\ process.} This \wsbpel\ process
can be experimented via the Web interface at
\url{http://reggae.dsi.unifi.it/blinddatejoining/}, or by downloading from
the same address the source and binary code.

More specifically, the process is configured to be deployed in a \wsbpel\
engine Apache ODE~\cite{ODE}. Indeed, we have equipped the process with the
corresponding \wsdl\ and deployment descriptor files, which provide typing
and binding information, respectively. Notably, in order to call the custom
XPath expression \code{getTableID()} within the process, its definition
must be previously installed in the ODE engine as a Java library. For the
sake of simplicity, we have defined such expression as a random function.

\begin{figure}[t]
  \begin{center}
    \includegraphics[scale=.37]{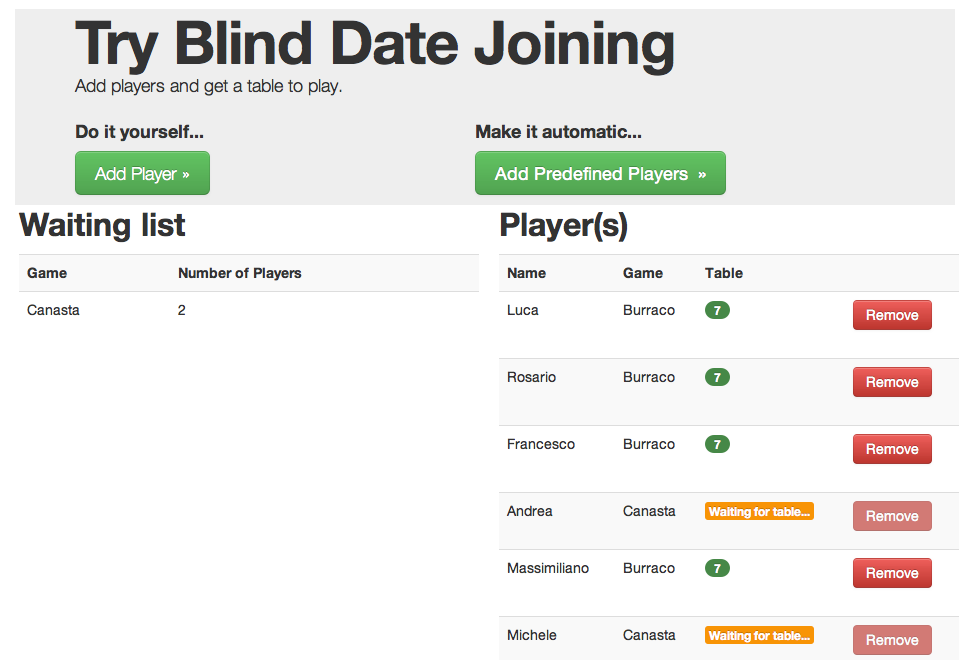}
  \end{center}
  \vspace*{-.7cm}
  \caption{A screenshot of the Web application for experimenting with the process \code{TableManager}}
  \label{fig:screenshot}
  \vspace*{-.3cm}
\end{figure}

To experiment with the \wsbpel\ program, we have developed a sort of
testing environment consisting of a few Java classes implementing the
service clients. Such classes rely on the artifacts automatically generated
by JAX-WS~\cite{JAX} from the \wsdl\ document of the process, and simply
exchange SOAP messages with the process. These clients are instantiated and
executed by a Web application developed by using the Play
framework~\cite{PLAY}; a screenshot of such application is shown in
Figure~\ref{fig:screenshot}. 
Notably, players created in a given browser session could be assigned to tables
together with players created in other sessions.

\section{Semantics mechanisms underlying blind-date conversation joining}
\label{sec:caseStudycows}

In this section, to clarify the behaviour at runtime of the
\code{TableManager} process (and of its instances), we formally specify a
scenario involving the manager service by means of the process calculus
\cows. The aim is to shed light on the effect of the blind-date joining and
to show how it can be easily programmed through the correlation approach.
Moreover, this formal account also permits clarifying the mechanisms
underlying message correlation (i.e., shared input variables,
pattern-matching and priority among receive activities).

The \code{TableManager} process can be rendered in \cows\ as follows:
$$
\begin{array}{r@{\,}c@{\,}l}
\mathit{TableManagerProcess} & \ \define \ &
\rep \scope{x_{game},x_{\mathit{player1}},x_{\mathit{player2}},x_{\mathit{player3}},x_{\mathit{player4}}}\\
&&\ \ \
\inp{\mathit{manager}}{\mathit{join}}{\arr{x_{\mathit{game}},x_{\mathit{player1}}}}.\\
&&\ \ \
\inp{\mathit{manager}}{\mathit{join}}{\arr{x_{\mathit{game}},x_{\mathit{player2}}}}.\\
&&\ \ \
\inp{\mathit{manager}}{\mathit{join}}{\arr{x_{\mathit{game}},x_{\mathit{player3}}}}.\\
&&\ \ \
\inp{\mathit{manager}}{\mathit{join}}{\arr{x_{\mathit{game}},x_{\mathit{player4}}}}.\\
&&\qquad
\scope{\mathit{tableId}}
(\, \out{x_{\mathit{player1}}}{\mathit{start}}{\arr{\mathit{tableId}}}\\
&&\qquad\qquad\qquad\ \ \,
    \spar \out{x_{\mathit{player2}}}{\mathit{start}}{\arr{\mathit{tableId}}}\\
&&\qquad\qquad\qquad\ \ \,
    \spar \out{x_{\mathit{player3}}}{\mathit{start}}{\arr{\mathit{tableId}}}\\
&&\qquad\qquad\qquad\ \ \,
    \spar \out{x_{\mathit{player4}}}{\mathit{start}}{\arr{\mathit{tableId}}}
\,)
\end{array}
$$
The replication operator $\rep$ specifies that the above term represents a
service definition, which acts as a persistent service capable of creating
multiple instances to simultaneously serve concurrent requests. The
delimitation operator $\scope{\ }$ declares the scope of variables
$x_{game}$ and $x_{\mathit{player\,i}}$, with $i=1..4$. The endpoint
\mbox{$\epc{\mathit{manager}}{\mathit{join}}$} (composed by the partner
name $\mathit{manager}$ and the operation $\mathit{join}$) is used by the
service to receive join requests from four players. When sending their
request, the players are required to provide only the kind of game, stored
in $x_{game}$, and their partner names, stored in $x_{\mathit{player\,i}}$,
that they will then use to receive the table identifier.  Player's requests
are received through receive activities of the form
$\inp{\mathit{manager}}{\mathit{join}}{\arr{x_{\mathit{game}},x_{\mathit{player\,i}}}}$,
which are correlated by means of the shared variable $x_{\mathit{game}}$.
Then, the delimitation operator is used to create a fresh name
$\mathit{tableId}$ that represents an unique table identifier. Such
identifier will be then communicated to each player by means of four invoke
activities
$\out{x_{\mathit{player\,i}}}{\mathit{start}}{\arr{\mathit{tableId}}}$.
Notably, differently from the \wsbpel\ specification of the process, in the
\cows\ definition the assign activities are not necessary, because their
role is played by the substitutions generated by the interactions along the
endpoint \mbox{$\epc{\mathit{manager}}{\mathit{join}}$}.

Consider now the following system
$$
\mathit{Luca}
\,\spar\,
\mathit{Rosario}
\,\spar\,
\mathit{Francesco}
\,\spar\,
\ldots
\,\spar\,
\mathit{TableManagerProcess}
$$
where the players are defined as follows
$$
\begin{array}{r@{\,}c@{\,}l}
\mathit{Luca} & \ \define \ &
\out{\mathit{manager}}{\mathit{join}}{\arr{\mathit{burraco},p_L}}
\spar \scope{x_{id}} \inp{p_L}{\mathit{start}}{\arr{x_{id}}}.\,
\arr{\mathrm{rest\ of}\ \mathit{Luca}}
\\[.4cm]
\mathit{Rosario} & \ \define \ &
\out{\mathit{manager}}{\mathit{join}}{\arr{\mathit{canasta},p_R}}
\spar \scope{x_{id}} \inp{p_R}{\mathit{start}}{\arr{x_{id}}}.\,
\arr{\mathrm{rest\ of}\ \mathit{Rosario}}
\\[.4cm]
\mathit{Francesco} & \ \define \ &
\out{\mathit{manager}}{\mathit{join}}{\arr{\mathit{burraco},p_F}}
\spar \scope{x_{id}} \inp{p_F}{\mathit{start}}{\arr{x_{id}}}.\,
\arr{\mathrm{rest\ of}\ \mathit{Francesco}}
\\[.2cm]
\multicolumn{1}{l}{\ldots}\\
\end{array}
$$

If $\mathit{Luca}$ requests to join a match, since there are not tables
under arrangement, $\mathit{TableManager}$-$\mathit{Process}$ initialises a
new match instance (highlighted by grey background) and the system evolves
to:
$$
\begin{array}{l}
\scope{x_{id}} \inp{p_L}{\mathit{start}}{\arr{x_{id}}}.\,
\arr{\mathrm{rest\ of}\ \mathit{Luca}}\\
\spar\,
\mathit{Rosario}
\,\spar\,
\mathit{Francesco}
\,\spar\, \ldots \,\spar\,
\mathit{TableManagerProcess}\\
\spar\,
\bgInstanceA{$
\scope{x_{\mathit{player2}},x_{\mathit{player3}},x_{\mathit{player4}}}$}\\
\ \ \
\bgInstanceA{$
\inp{\mathit{manager}}{\mathit{join}}{\arr{{\mathit{burraco}},x_{\mathit{player2}}}}.$}\\
\ \ \
\bgInstanceA{$
\inp{\mathit{manager}}{\mathit{join}}{\arr{{\mathit{burraco}},x_{\mathit{player3}}}}.$}\\
\ \ \
\bgInstanceA{$
\inp{\mathit{manager}}{\mathit{join}}{\arr{{\mathit{burraco}},x_{\mathit{player4}}}}.$}\\
\qquad
\bgInstanceA{$
\scope{\mathit{tableId}}
(\, \out{p_L}{\mathit{start}}{\arr{\mathit{tableId}}}$}\\
\qquad\qquad\qquad\ \ \,
\bgInstanceA{$
    \spar \out{x_{\mathit{player2}}}{\mathit{start}}{\arr{\mathit{tableId}}}$}\\
\qquad\qquad\qquad\ \ \,
\bgInstanceA{$
    \spar \out{x_{\mathit{player3}}}{\mathit{start}}{\arr{\mathit{tableId}}}$}\\
\qquad\qquad\qquad\ \ \,
\bgInstanceA{$
    \spar \out{x_{\mathit{player4}}}{\mathit{start}}{\arr{\mathit{tableId}}}
\,)$}
\end{array}
$$
Now, if $\mathit{Rosario}$ invokes $\mathit{TableManagerProcess}$, a second
match instance (highlighted by dark grey background)  is created:
$$
\begin{array}{l}
\scope{x_{id}} \inp{p_L}{\mathit{start}}{\arr{x_{id}}}.\,
\arr{\mathrm{rest\ of}\ \mathit{Luca}}\\
\spar\,
\scope{x_{id}} \inp{p_R}{\mathit{start}}{\arr{x_{id}}}.\,
\arr{\mathrm{rest\ of}\ \mathit{Rosario}}\\
\spar\,
\mathit{Francesco}
\,\spar\, \ldots \,\spar\,
\mathit{TableManagerProcess}\\
\spar\,
\bgInstanceA{$
\scope{x_{\mathit{player2}},x_{\mathit{player3}},x_{\mathit{player4}}}$}\\
\ \ \
\bgInstanceA{$
\inp{\mathit{manager}}{\mathit{join}}{\arr{{\mathit{burraco}},x_{\mathit{player2}}}}.$}\\
\ \ \
\bgInstanceA{$
\inp{\mathit{manager}}{\mathit{join}}{\arr{{\mathit{burraco}},x_{\mathit{player3}}}}.$}\\
\ \ \
\bgInstanceA{$
\inp{\mathit{manager}}{\mathit{join}}{\arr{{\mathit{burraco}},x_{\mathit{player4}}}}.$}\\
\qquad
\bgInstanceA{$
\scope{\mathit{tableId}}
(\, \out{p_L}{\mathit{start}}{\arr{\mathit{tableId}}}$}\\
\qquad\qquad\qquad\ \ \,
\bgInstanceA{$
    \spar \out{x_{\mathit{player2}}}{\mathit{start}}{\arr{\mathit{tableId}}}$}\\
\qquad\qquad\qquad\ \ \,
\bgInstanceA{$
    \spar \out{x_{\mathit{player3}}}{\mathit{start}}{\arr{\mathit{tableId}}}$}\\
\qquad\qquad\qquad\ \ \,
\bgInstanceA{$
    \spar \out{x_{\mathit{player4}}}{\mathit{start}}{\arr{\mathit{tableId}}}
\,)$}
\end{array}
$$
$$
\begin{array}{l}
\spar\,
\bgInstanceB{$
\scope{x_{\mathit{player2}},x_{\mathit{player3}},x_{\mathit{player4}}}$}\\
\ \ \
\bgInstanceB{$
\inp{\mathit{manager}}{\mathit{join}}{\arr{{\mathit{canasta}},x_{\mathit{player2}}}}.$}\\
\ \ \
\bgInstanceB{$
\inp{\mathit{manager}}{\mathit{join}}{\arr{{\mathit{canasta}},x_{\mathit{player3}}}}.$}\\
\ \ \
\bgInstanceB{$
\inp{\mathit{manager}}{\mathit{join}}{\arr{{\mathit{canasta}},x_{\mathit{player4}}}}.$}\\
\qquad
\bgInstanceB{$
\scope{\mathit{tableId}'}
(\, \out{p_R}{\mathit{start}}{\arr{\mathit{tableId}'}}$}\\
\qquad\qquad\qquad\ \ \ \,
\bgInstanceB{$
    \spar \out{x_{\mathit{player2}}}{\mathit{start}}{\arr{\mathit{tableId}'}}$}\\
\qquad\qquad\qquad\ \ \ \,
\bgInstanceB{$
    \spar \out{x_{\mathit{player3}}}{\mathit{start}}{\arr{\mathit{tableId}'}}$}\\
\qquad\qquad\qquad\ \ \ \,
\bgInstanceB{$
    \spar \out{x_{\mathit{player4}}}{\mathit{start}}{\arr{\mathit{tableId}'}}
\,)$}
\end{array}
$$

When $\mathit{Francesco}$ invokes $\mathit{TableManagerProcess}$, the
process definition and the first created instance, being both able to
receive the same message $\arr{\mathit{burraco},p_F}$ along the endpoint
\mbox{$\epc{\mathit{manager}}{\mathit{join}}$}, compete for the request
$\out{\mathit{manager}}{\mathit{join}}{\arr{\mathit{burraco},p_F}}$.  \cows
's (prioritized) semantics precisely establishes how this sort of race
condition is dealt with: only the existing instance is allowed to evolve,
as required by \wsbpel. This is done through the dynamic prioritised
mechanism of \cows, i.e. assigning the receives performed by instances
(having a more defined pattern and requiring less substitutions) a greater
priority than the receives performed by a process definition. In fact, in
the above \cows\ term, the first instance can perform a receive matching
the message and containing only one variable in its argument, while the
initial receive of $\mathit{TableManagerProcess}$ contains two variables.
In this way, the creation of a new instance is prevented. Moreover,
pattern-matching permits delivering the request to the appropriate
instance, i.e. that corresponding to a $\mathit{burraco}$ match. Therefore,
the only feasible computation leads to the following term
$$
\begin{array}{l}
\scope{x_{id}} \inp{p_L}{\mathit{start}}{\arr{x_{id}}}.\,
\arr{\mathrm{rest\ of}\ \mathit{Luca}}\\
\spar\, \scope{x_{id}} \inp{p_R}{\mathit{start}}{\arr{x_{id}}}.\,
\arr{\mathrm{rest\ of}\ \mathit{Rosario}}\\
\spar\, \scope{x_{id}} \inp{p_F}{\mathit{start}}{\arr{x_{id}}}.\,
\arr{\mathrm{rest\ of}\ \mathit{Francesco}}\\
\spar\, \ldots \spar\,
\mathit{TableManagerProcess}\\
\spar\,
\bgInstanceA{$
\scope{x_{\mathit{player3}},x_{\mathit{player4}}}$}\\
\ \ \
\bgInstanceA{$
\inp{\mathit{manager}}{\mathit{join}}{\arr{{\mathit{burraco}},x_{\mathit{player3}}}}.$}\\
\ \ \
\bgInstanceA{$
\inp{\mathit{manager}}{\mathit{join}}{\arr{{\mathit{burraco}},x_{\mathit{player4}}}}.$}\\
\qquad
\bgInstanceA{$
\scope{\mathit{tableId}}
(\, \out{p_L}{\mathit{start}}{\arr{\mathit{tableId}}}$}\\
\qquad\qquad\qquad\ \ \,
\bgInstanceA{$
    \spar \out{p_F}{\mathit{start}}{\arr{\mathit{tableId}}}$}\\
\qquad\qquad\qquad\ \ \,
\bgInstanceA{$
    \spar \out{x_{\mathit{player3}}}{\mathit{start}}{\arr{\mathit{tableId}}}$}\\
\qquad\qquad\qquad\ \ \,
\bgInstanceA{$
    \spar \out{x_{\mathit{player4}}}{\mathit{start}}{\arr{\mathit{tableId}}}
\,)$}
\\
\spar\,
\bgInstanceB{$
\scope{x_{\mathit{player2}},x_{\mathit{player3}},x_{\mathit{player4}}}$}\\
\ \ \
\bgInstanceB{$
\inp{\mathit{manager}}{\mathit{join}}{\arr{{\mathit{canasta}},x_{\mathit{player2}}}}.$}\\
\ \ \
\bgInstanceB{$
\inp{\mathit{manager}}{\mathit{join}}{\arr{{\mathit{canasta}},x_{\mathit{player3}}}}.$}\\
\ \ \
\bgInstanceB{$
\inp{\mathit{manager}}{\mathit{join}}{\arr{{\mathit{canasta}},x_{\mathit{player4}}}}.$}\\
\qquad
\bgInstanceB{$
\scope{\mathit{tableId}'}
(\, \out{p_R}{\mathit{start}}{\arr{\mathit{tableId}'}}$}\\
\qquad\qquad\qquad\ \ \ \,
\bgInstanceB{$
    \spar \out{x_{\mathit{player2}}}{\mathit{start}}{\arr{\mathit{tableId}'}}$}\\
\qquad\qquad\qquad\ \ \ \,
\bgInstanceB{$
    \spar \out{x_{\mathit{player3}}}{\mathit{start}}{\arr{\mathit{tableId}'}}$}\\
\qquad\qquad\qquad\ \ \ \,
\bgInstanceB{$
    \spar \out{x_{\mathit{player4}}}{\mathit{start}}{\arr{\mathit{tableId}'}}
\,)$}
\end{array}
$$
where $\mathit{Francesco}$ joined to the $\mathit{burraco}$ table under
arrangement.

Eventually, with the arrival of other requests from players that want to
play $\mathit{burraco}$, the $\mathit{TableManagerProcess}$ completes to
arrange the $\mathit{burraco}$ table and contacts the players for
communicating them the table identifier:
$$
\begin{array}{l}
\scope{\mathit{tableId}}
(\,\arr{\mathrm{rest\ of}\ \mathit{Luca}}
\,\spar\, \arr{\mathrm{rest\ of}\ \mathit{Francesco}} \,\spar\, \ldots\,)\\
\spar\, \scope{x_{id}} \inp{p_R}{\mathit{start}}{\arr{x_{id}}}.\,
\arr{\mathrm{rest\ of}\ \mathit{Rosario}}\\
\spar\, \ldots\spar\,
\mathit{TableManagerProcess}\\
\spar\,
\bgInstanceB{$
\scope{x_{\mathit{player2}},x_{\mathit{player3}},x_{\mathit{player4}}}$}\\
\ \ \
\bgInstanceB{$
\inp{\mathit{manager}}{\mathit{join}}{\arr{{\mathit{canasta}},x_{\mathit{player2}}}}.$}\\
\ \ \
\bgInstanceB{$
\inp{\mathit{manager}}{\mathit{join}}{\arr{{\mathit{canasta}},x_{\mathit{player3}}}}.$}\\
\ \ \
\bgInstanceB{$
\inp{\mathit{manager}}{\mathit{join}}{\arr{{\mathit{canasta}},x_{\mathit{player4}}}}.$}\\
\qquad
\bgInstanceB{$
\scope{\mathit{tableId}'}
(\, \out{p_R}{\mathit{start}}{\arr{\mathit{tableId}'}}$}\\
\qquad\qquad\qquad\ \ \ \,
\bgInstanceB{$
    \spar \out{x_{\mathit{player2}}}{\mathit{start}}{\arr{\mathit{tableId}'}}$}\\
\qquad\qquad\qquad\ \ \ \,
\bgInstanceB{$
    \spar \out{x_{\mathit{player3}}}{\mathit{start}}{\arr{\mathit{tableId}'}}$}\\
\qquad\qquad\qquad\ \ \ \,
\bgInstanceB{$
    \spar \out{x_{\mathit{player4}}}{\mathit{start}}{\arr{\mathit{tableId}'}}
\,)$}\end{array}
$$
Therefore, the players of table $\mathit{tableId}$ (including
$\mathit{Luca}$ and $\mathit{Francesco}$) can start to play, while
$\mathit{Rosario}$ continues to wait for other $\mathit{canasta}$ players.


\section{Concluding remarks}
\label{sec:conclusion}

We have illustrated blind-date conversation joining, a strategy allowing a
participant to join a conversation without need to know information about
the conversation itself, such as e.g. its identifier or the other
participants, but the endpoint of the service provider. This is possible
because some values in the request define which conversation the
participant will join. In case such a conversation does not exist, a new
one associated to these values will be created.

We showed how the blind-date conversation joining strategy can be
implemented in the well-established industrial standard \wsbpel. In order
to accomplish this task, we used \wsbpel\ correlation sets to filter the
client requests and create conversations identified by means of the
filtered values.

We also presented a formal semantics for the strategy by using the process
calculus \cows. The semantics permits to highlight what the effect of the
strategy is, by clarifying the mechanism underlying message correlation and
showing how the strategy can be easily programmed in a correlation-based
environment. This also permits appreciating the conciseness and cleanness
of the \cows\ specification of the \code{TableManager} process, especially
when compared with its \wsbpel\ counterpart.

We developed a case study to show how our strategy can be used in a
realistic scenario. The case study defines an online games provider that
arranges online matches of 4-players card games. This scenario is
implemented using \wsbpel\ and executed on Apache ODE. As we showed in
Section~\ref{sec:wsbpel}, despite the XML markup used by \wsbpel, the
blind-date joining strategy permits to have a smooth and clean
implementation. We also equipped the case study with a `testing
environment' in order to facilitate its testing.

\paragraph{Related work.}
\label{sec:relWork}

The peculiar form of conversation joining studied in this paper, which we
call `blind-date', originates from the message correlation mechanism used
for delivering  messages to the appropriate service instances in both
orchestration languages and formalisms for SOC. This joining strategy is,
at least in principle, independent from the specific language or formalism
used to enact it. In this paper, we have used the language \wsbpel\ and the
formalism \cows, but different choices could have been made. For example,
Jolie~\cite{jolie} could be used as correlation-based orchestration
language and SOCK~\cite{SOCK} as the formalism, being the former a
Java-based implementation of the latter. However, our choice fell on
\wsbpel\ because it is an OASIS open standard well-accepted by industries
and, hence, supported by well-established and maintained engines. Instead,
\cows\ has been selected because of its strict correspondence with \wsbpel\
while, at the same time, being a core calculus consisting of just a few
constructs, which makes it more suitable than \wsbpel\ to reason on
applications' behaviour.

In the SOC literature, two main approaches have been considered to connect
the interaction protocols of clients and of the respective service
instances. That based on the correlation mechanism was first exploited in
\cite{Viroli} where, however, only interaction among different instances of
a single service are taken into account. Another correlation-based
formalism, besides \cows\ and SOCK mentioned above, is the calculus
\emph{Corr}~\cite{MelgrattiR11}, which is a sort of value-passing CCS
without restriction and enriched with constructs for expressing services
and their instances. \emph{Corr} has been specifically designed to capture
behaviours related to correlation aspects in a specific \wsbpel\ engine.
Another work with an aim similar to ours, i.e. to show an exploitation of
the correlation-based mechanism for dealing with issues raised by practical
scenarios, is presented in \cite{MauroGGM11}. Such work proposes an
implementation of a correlation-based primitive allowing messages to be
delivered to more than one service conversation. Anyway, \wsbpel\ (and
\cows\ as well) natively provides such kind of broadcast primitive.

A large strand of work, instead, relies on the explicit modelling of
interaction \emph{sessions} and their dynamic creation. A session
corresponds to a private channel (\emph{\`a la} \pic~\cite{PICALC}) which
is implicitly instantiated when calling a service: it binds caller and
callee and is used for their future conversation.
Although this the technology underling SOC, its abstraction level has
proved convenient for reasoning about SOC applications. Indeed,
session-based conversation can be regulated by so called \emph{session
types} that can statically guarantee a number of desirable properties, such
as communication safety, progress, predictability.  scenario declared in
the session type).  Therefore, an important group of calculi for modeling
and proving properties of services is based on the explicit notion of
interaction session. Most of such work (among which we mention
\cite{SessionTypesHVK98,carboneesop,LMRV07,Caspis}) has been devoted to
study \emph{dyadic} sessions, i.e. interaction sessions between only two
participants.

Recently, another body of work (as, e.g.,
\cite{MultipartyAsyncST,Muse,CC,CarboneM13}) focussed on a more general
form of sessions, called \emph{multiparty} sessions/conversations, which
are closer to the notion of conversation that we have considered in this
paper. The multiparty session approach proposed in~\cite{MultipartyAsyncST}
permits expressing a conversation by means of channels shared among the
participants. However, the conversation is created through a single
synchronization among all participants, whose number is fixed at design
time. This differs from our notion of blind-date joining, where once a
conversation has been created the participants can asynchronously join.
Moreover, although in our case study we consider a conversation of five
participants, in principle this number can change at runtime. The $\mu$se
language~\cite{Muse} permits the declaration of multiparty sessions in a
way transparent to the user. However, rather than relying on the
correlation mechanism, as in our approach, $\mu$se creates multiparty
conversations by using a specific primitive that permits to merge together
previously created conversations. Instead, when relying on correlation, the
conversation merging is automatically performed for each correlated
request. Another formalism dealing with multiparty interactions is the
Conversation Calculus~\cite{VieiraCS08,CC}. It uses the conversation-based
mechanism for inter-session communication, which permits to progressively
accommodate and dismiss participants from the same conversation. This is
realized by means of named containers for processes, called conversation
contexts. However, processes in unrelated conversations cannot interact
directly. Moreover, conversation joining is not transparent to a new
participant, because he has to know the name of the conversation.  Instead,
in our correlation-based approach, a conversation is represented by a
service instance, which is not accessed via an identifier, but via the
correlation values specified by the correlation set.

In conclusion, the lower-level mechanism based on correlation sets, that
exploits business data to correlate different interactions, is more robust
and fits the loosely-coupled world of Web Services better than that based
on explicit session references. It turned out to be powerful and flexible
enough to fulfill the need of SOC, e.g. it easily allows a single message
to participate in multiple conversations, each (possibly) identified by
separate correlation values, which instead requires non-trivial workarounds
by using the session-based approach. It is not a case that also the
standard \wsbpel\ uses correlation sets.

\paragraph{Future work.}

To better highlight the semantics mechanisms underlying blind-date conversation joining, we have preferred to consider a quite simple case study. Anyway, many other features could be added without much effort that would make the case study even more realistic. For example, a player should be able to play concurrently in more than one match. However, for the sake of simplicity, we assume here that a player can ask to play the same game more than once only if each time it waits for the response to the previous request, otherwise it could be assigned more than once to the same table.
Moreover, $\mathit{TableManagerProcess}$ can be easily tailored to arrange matches for $n$-players online games with $n\neq 4$ (e.g. poker, bridge, \ldots). It could also be modified so to be able to manage games with a number of players not statically known, as well as games that can be played by a variable number of players, i.e.~games that may begin with a minimum number of players but later on other players can dynamically join and some others can stop playing and leave.
The fact that some players may give up, i.e. the number of players may also decrease, implies that some tables may dynamically have not enough players for the game to go on. In this case, tables with an insufficient number of players could be joined and there are several possible strategies to to this: a centralized solution could exploit the table manager as a forwarder for the player messages, while a decentralized one could exploit the tables with not enough players as forwarders; alternatively, we could exploit \emph{memory cell} services to store the (dynamically updatable) value of the table assigned to each single player thus avoiding to replace, once and for all, the correlation variable with this value; finally, $\mathit{TableManagerProcess}$  instances, instead of single tables, could represent rooms containing more tables for the same game and manage all the requests for that game.
In order to avoid deadlocks, we could add timeouts to receive activities of $\mathit{TableManagerProcess}$ and players' services. This can be done both in \wsbpel, by exploiting \code{onAlarm} events within \code{pick} activities, and in \cows , by using \emph{wait} activities in conjunction with choice activities (as shown in \cite{COWS_JAL}).

We leave as a future work the task of formalizing the pattern underlying blind-date conversation joining in a way which is independently from the specific case study considered in this paper. We also plan to study and develop type theories for describing and regulating the correlation mechanism. This will permit to guarantee desired properties like those defined for dyadic and multiparty session types. We expect that to develop type theories for ensuring properties of the correlation mechanism be more difficult than doing so for session-based interaction because the communication media are not necessarily private and the correlation values are determined at runtime. 

\bibliographystyle{eptcs}
\bibliography{biblio}

\end{document}